\documentstyle[twoside,fleqn,npb,epsfig]{article}
\def\ppg{\pi^{+}\pi^{-}\gamma}
\def\mmg{\mu^{+}\mu^{-}\gamma}
\def\eeg{e^{+}e^{-}\gamma}
\def\ie{{\it\kern-2pt i.\kern-.5pt e.\kern-2pt}}  
\def\ifm#1{\relax\ifmmode#1\else$#1$\fi}
     
\def\epm{\ifm{e^+e^-}} \def\gam{\ifm{\gamma}}
  
\def\to{\ifm{\rightarrow}} \def\sig{\ifm{\sigma}}  
\def\up#1{\ifm{^{#1}}}  
  \def\dif{\hbox{d}}  
   \def\x{\ifm{\times}}
  \def\pt#1,#2,{\ifm{#1\x10^{#2}}}
   
\newcommand{\AmS}{{\protect\the\textfont2
  A\kern-.1667em\lower.5ex\hbox{M}\kern-.125emS}}

\title{Measuring the Hadronic Cross Section via Radiative Return}

\author{The KLOE Collaboration~\thanks{
The KLOE Collaboration: A.~Aloisio,
F.~Ambrosino,
A.~Antonelli,
M.~Antonelli,
C.~Bacci,
G.~Bencivenni,
S.~Bertolucci,
C.~Bini,
C.~Bloise,
V.~Bocci,
F.~Bossi,
P.~Branchini,
S.~A.~Bulychjov,
R.~Caloi,
P.~Campana,
G.~Capon,
G.~Carboni,
M.~Casarsa,
V.~Casavola,
G.~Cataldi,
F.~Ceradini,
F.~Cervelli,
F.~Cevenini,
G.~Chiefari,
P.~Ciambrone,
S.~Conetti,
E.~De~Lucia,
G.~De~Robertis,
P.~De~Simone,
G.~De~Zorzi,
S.~Dell'Agnello,
A.~Denig,
A.~Di~Domenico,
C.~Di~Donato,
S.~Di~Falco,
A.~Doria,
M.~Dreucci,
O.~Erriquez,
A.~Farilla,
G.~Felici,
A.~Ferrari,
M.~L.~Ferrer,
G.~Finocchiaro,
C.~Forti,
A.~Franceschi,
P.~Franzini,
C.~Gatti,
P.~Gauzzi,
S.~Giovannella,
E.~Gorini,
F.~Grancagnolo,
E.~Graziani,
S.~W.~Han,
M.~Incagli,
L.~Ingrosso,
W.~Kluge,
C.~Kuo,
V.~Kulikov,
F.~Lacava,
G.~Lanfranchi,
J.~Lee-Franzini,
D.~Leone,
F.~Lu,
M.~Martemianov,
M.~Matsyuk,
W.~Mei,
L.~Merola,
R.~Messi,
S.~Miscetti,
M.~Moulson,
S.~M\"uller,
F.~Murtas,
M.~Napolitano,
A.~Nedosekin,
F.~Nguyen,
M.~Palutan,
L.~Paoluzi,
E.~Pasqualucci,
L.~Passalacqua,
A.~Passeri,
V.~Patera,
E.~Petrolo,
L.~Pontecorvo,
M.~Primavera,
F.~Ruggieri,
P.~Santangelo,
E.~Santovetti,
G.~Saracino,
R.~D.~Schamberger,
B.~Sciascia,
A.~Sciubba,
F.~Scuri,
I.~Sfiligoi,
T.~Spadaro,
E.~Spiriti,
G.~L.~Tong,
L.~Tortora,
E.~Valente,
P.~Valente,
B.~Valeriani,
G.~Venanzoni,
S.~Veneziano,
A.~Ventura,
G.~Xu,
G.~W.~Yu.
} 
\\presented by Achim G. Denig
\\
{\it IEKP-Universit\"at Karlsruhe, Postfach 3640, 76021 Karlsruhe, 
Germany}}

\begin{document}

\begin{abstract}\noindent
Recently it has been demonstrated that particle factories, such as  DA$\Phi$NE and PEP-II, operating at fixed center-of-mass energies, are able to measure hadronic cross sections as a function of the hadronic system energy using the radiative return. This paper is an experimental overview of the progress in this area. Preliminary results from KLOE for the process $e^{+}e^{-} \rightarrow \rho \gamma  \rightarrow \pi^{+}\pi^{-}\gamma$ and a fit to the pion form factor are presented. Some first results from the BABAR collaboration are also shown. 
\end{abstract}

\maketitle

\section{Introduction}

\subsection{Motivation}
A complete estimate of the muon anomaly $a_\mu = (g_\mu-2)/2$, where $g_\mu$ is the muon gyromagnetic ratio, requires computation of the hadronic contributions to the photon propagator spectral function. Perturbative QCD fails at low energy. The vacuum polarization correction to the photon propagator is closely related to the hadronic cross section. The hadronic correction to the muon anomaly $a_\mu({\rm hadr})$ can therefore be calculated from measurements of the hadronic cross section in \epm\ annihilations via a dispersion relation:
\begin{equation}
a_\mu({\rm hadr})={1\over4\pi^3}\int_{4 m_{\pi}^2}^{\infty}%
\sigma_{\epm\to{\rm hadr}}(s)K(s)\dif s.
\label{amu}
\end{equation}
The kernel $K(s)$ is given by many authors and behaves approximately as $1/s$. The annihilation cross section also has an intrinsic $1/s$ dependence and is largely enhanced around $M(\pi\pi)\sim$770 MeV by the $\rho$ pole. Data at low energies therefore contribute strongly to $a_\mu({\rm hadr})$.
The present uncertainty of the hadronic contribution to the anomaly is dominated by the limited accuracy of hadronic cross section and $\tau$ data.

For a detailed discussion and the interpretation of possible discrepancies between the theoretical and the experimental value for $a_\mu$: \\
$a_\mu^{theo} = (11659186.3 \pm 7.1)$x$10^{-10}$ [$\tau$ data]\\
$a_\mu^{theo} = (11659169.1 \pm 7.8)$x$10^{-10}$ [$e^+ e^-$ data]\\
$a_\mu^{exp} =  (11659203 \pm 8)$x$10^{-10}$ [E821],\\
see refs.~\cite{Davier:2002dy}~\cite{Czarnecki:2001pv}.

The first $a_\mu$ value above is obtained including $\tau$ decay data, assuming conservation of the vector current and isospin symmetry~\cite{Davier:1998si}.
The second value uses only $e^+e^-$ data (see also~\cite{EidJeg95}), in particular the new results from Novosibirsk (CMD-2~\cite{Akhmetshin:2001ig}, $\sqrt{s}<1.4$ GeV) and Beijing 
(BES~\cite{bes}, $2\ {\rm GeV}<\sqrt{s}<5$ GeV). 
The new CMD-2 measurement of the dominant
process $e^+ e^- \rightarrow \rho \rightarrow \pi^+ \pi^-$
with a $0.6\%$ precision is chiefly responsible for the signi\-fi\-cant reduction of the error of $a_\mu^{hadr}$ in the second analysis, which now shows a $3.0 \sigma$ deviation from the experimental value, while the analysis based on $\tau$ data gives only a $1.6 \sigma$ effect. Clearly, new, precise hadronic cross section data are needed to cross
check the current discrepancy between theory and experiment and 
to clarify the difference between the two analyses.

\subsection{Radiative Return}
We present in this paper a method to obtain $\sig(\epm\to{\rm hadrons})$, which employs the radiative process \epm\to hadrons+\gam, where the photon has been radiated by one of the initial electrons or positrons, Initial State Radiation, ISR~\cite{Spagnolo:1998mt}~\cite{Binner:1999bt}. 
Particle factories such as DA$\Phi$NE and PEP-II, typically operate at fixed center-of-mass energies: $W=m_\phi$ at DA$\Phi$NE, $W=m_{\Upsilon(4S)}$ at PEP-II.
By measuring the radiative return process above, the hadronic cross section becomes accessible over the mass range $M_{\rm hadr} < W$. The cross section $\dif\sigma(\epm\to{\rm hadrons})/\dif s$ can be obtained from the measurement of $\dif\sigma(\epm\to{\rm hadrons}+\gam)/\dif s'$, $s'=M_{\rm hadr}^2$. 
The relation is given by the radiation function $H$ defined by:
\begin{equation}
s\:{\dif\sigma({\rm hadrons}+\gamma)\over\dif s'} =
\sigma({\rm hadrons}) \x H(s,\theta_\gamma) 
\label{H}
\end{equation}

The function $H(s,\theta_\gamma)$ (which depends also on the polar angle of the photon $\theta_\gamma$) needs to be known to an accuracy better than $1\%$ for precision measurements. Radiative corrections have been computed, so far, by different groups~\cite{czyz02}~\cite{Hoefer:2001mx}~\cite{Khoze:2002ix}
up to next-to-leading-order, for the exclusive final hadronic 
states $\pi^+ \pi^- \gamma$ 
and 4$\pi+\gamma$~\cite{czyz01}.
Our KLOE analysis is based on the yielding theoretical 
event generators EVA~\cite{Binner:1999bt} and PHOKHARA~\cite{Rodrigo:2001jr}~\cite{Rodrigo:2001kf}~\cite{Kuhn:2002xg}.
\\
We present in the following preliminary KLOE results for the process
$e^+ e^- \to \rho \gamma \to \pi^+ \pi^- \gamma$, which
dominates the cross section below 1 GeV. This channel accounts for 62$\%$ of the total hadronic contribution to $a_\mu$.

Preliminary results of the BABAR collaboration (PEP-II collider) 
are presented at the end of this paper, where measurements using radiative return are performed at higher energies. At PEP-II one can in principle cover the whole energy range of interest for the evaluation of $a_\mu$. Moreover, measurements at higher energies are very important for evaluating the hadronic contribution to the running fine structure constant $\alpha(M_{Z})$ (hadronic vacuum polarization), which is given by a similar dispersion relation as the one shown for $a_\mu$. 

\section{KLOE analysis $e^{+}e^{-} \rightarrow \ppg$}

\subsection{Signal selection and backgound}

The selection of $e^{+}e^{-} \rightarrow \ppg$ events with the 
KLOE detector is presented in the following. 
A detailed description of the KLOE detector, which consists of a 
high resolution tracking detector ($\sigma_{p_T} / p_T \leq 0.3\%$) 
and an electromagnetic calorimeter ($\sigma_E / E = 5.7 \% /  \sqrt{E({\rm GeV})}$) 
can be found in references~\cite{dc}~\cite{emc}.
In this paper we describe an analysis where
the hard photon of the $\ppg$ events is emitted with small polar 
angles ($\theta_\gamma<15^\circ$). No explicit photon detection is done since the electromagnetic calorimeter has low acceptance at 
small angles. But as we will show in the following, that 
an efficient and almost background free signal selection can be 
obtained in this case without photon tagging~\cite{Cataldi:1999dc}~\cite{lp01}~\cite{achim}. 

\begin{itemize} 
\item {\it Detection of two charged tracks}: with polar angle larger than $40^\circ$, coming from a vertex in the fiducial volume $R<8$ cm, $|z|< 15$ cm. The cuts on the transverse momentum $p_{T} > 200$ MeV or on the longitudinal momentum $|p_{z}| > 90$ MeV reject tracks spiralizing along the beam line, ensuring good reconstruction conditions. 
The probability to reconstruct a vertex in the drift chamber is $\sim 98\%$ and 
has been studied with $\pi^+ \pi^- \pi^0$ and $\pi^+ \pi^-$ data.
%
\item {\it Identification of pion tracks}: A Likelihood Method 
(calibrated on real data), using the time of flight of the particle and the shape of the energy deposit in the electromagnetic calorimeter, 
has been developed to reject $e^{+}e^{-} \rightarrow \eeg$ background. 
$\eeg$ events are thus drastically reduced. A control sample of 
$\pi^{+}\pi^{-}\pi^{0}$ has been used to study the behaviour of 
pions in the electromagnetic calorimeter and to evaluate the selection 
efficiency for signal events, it is larger than $98\%$. 
%
%
%
\item {\it Cut on the track mass}: $\mmg$ events are rejected by a 
cut at 120 MeV in a kinematic variable called track mass.  
This variable is calculated from the reconstructed momenta, 
$\vec{p}_{+}$, $\vec{p}_{-}$, applying 4-momentum conservation under the hypothesis that the final state consists of two particles 
with the same mass and one photon. 
After this cut we find a contamination of $\mmg$ background smaller than $1\%$. 
\\
$\pi^{+}\pi^{-}\pi^{0}$ events are rejected with a cut in the 
two-dimensional distribution of the track mass versus the two pion 
invariant mass squared, $M_{\pi\pi}^{2}$.
The residual conta\-mi\-na\-tion of $\pi^{+}\pi^{-}\pi^{0}$ events
is expected to be at small $M_{\pi\pi}^{2}$ values ($M_{\pi\pi}^{2} < 0.4-0.5$ GeV$^{2}$). The efficiency of the track mass cut, as evaluated from MC, is $\sim 90\%$.
\item {\it Definition of the angular acceptance}: 
The di-pion production angle, $\theta_{\pi\pi}$, 
is calculated from the momenta of the two pions. If only one photon is emitted, the photon polar angle $\theta_{\gamma}=180^\circ-\theta_{\pi\pi}$. 
We select events with $\theta_{\pi\pi}<15^\circ$.\footnote{
A complementary analysis in which photons are selected at large
angles ($55^\circ<\theta_{\pi\pi}<125^\circ$) is in progress.
In this case the photon can be tagged in the electromagnetic 
calorimeter and the kinematical acceptance allows us to measure 
events down to the $2\pi$ threshold, in contrast to the small photon angles case where the acceptance vanishes for $s'<$0.24 GeV\up2.}

\end{itemize}

\begin{figure}[t] 
\vspace{6pt}
\includegraphics[width=17pc]{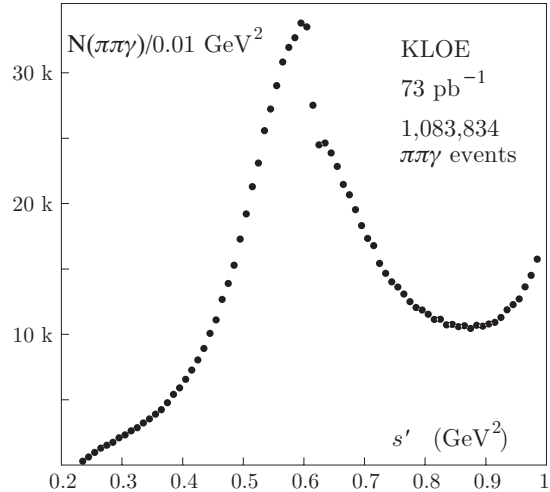}\vglue-.7cm
\caption{Number of $\ppg$ events selected by the analysis in the region 
$\theta_{\gamma}<15^\circ$ or $\theta_{\gamma}>165^\circ$, $40^\circ<\theta_{\pi}<140^\circ$.}
\label{fig:kloe}
\end{figure}

\subsection{Final state radiation}
An important background to our signal are events where the hard photon has been emitted by one of the final state pions, FSR. 
Since an initial state photon is emitted preferably at small polar 
angles, and the polar angle distribution of the photon from FSR 
follows the pion sin$^{2}\theta_{\pi}$ distribution, 
the FSR to ISR ratio changes con\-si\-derably with the polar 
angle of the two pion system, $\theta_{\pi\pi}$. For the small angle
phase space region described above ($\theta_{\pi\pi} < 15^\circ$) the
contribution of FSR is well below $1\%$ in all the 
$M_{\pi\pi}$ region.

Note that FSR events - as described in the case of the radiative return -  must not be included into the dispersion integral for $a_{\mu}$. In contrast, in the case of an energy scan, FSR should be included (see Ref.~\cite{Davier:2002dy}~\cite{Akhmetshin:2001ig}). 
In the case of the radiative return, FSR is emitted from a fixed value $M_{\pi\pi} = W$ (the beam center-of-mass energy), while for the dispersion integral, FSR is needed for the individual values  of $M_{\pi\pi}$ (as measured in a scan). 

\section{KLOE preliminary results}

\subsection{Effective cross section}
The $\pi^+ \pi^- \gamma$ cross section measurement
contains the following terms which are discussed
in the following:
\begin{equation}
\frac{d\sigma_{{\rm hadrons}+\gamma}}{dM_{\pi\pi}^2}=
\frac{dN_{{\rm Obs}}-dN_{{\rm Bkg}}}{dM_{\pi\pi}^2} \: 
\frac{1}{\epsilon_{\rm Sel}} \:
\frac{1}{\int{\mathcal{L}}dt}
\end{equation}

\begin{figure}[t]
\vspace{6pt}
\includegraphics[width=17pc]{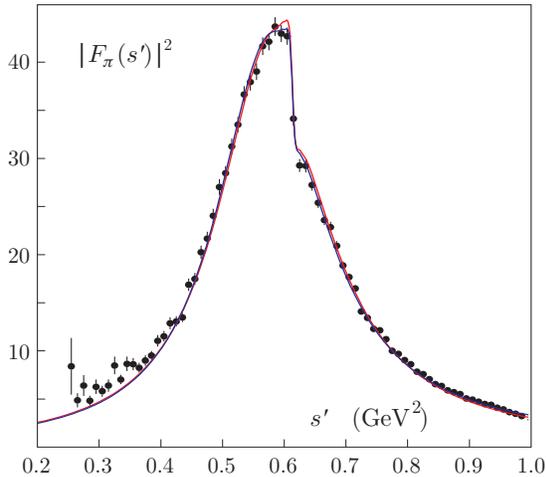}\vglue-.7cm
\caption{Preliminary measurement of the pion form factor. Superimposed are the analytical  $|F_\pi(s')|^2$ parametrizations, obtained by a fit to the KLOE data using the K\"uhn-Santamaria parametrization ({\it blue line}) and the parametrization found by CMD-2 with the Gounaris-Sakurai model  ({\it red line}). The two curves are in very good agreement.}
\label{fig:pion}
\end{figure}

\begin{itemize} 
\item{ Residual background $N_{Bkg}$ after event selection has to be 
studied (see 2.2). }
\item{ The selection efficiency $\epsilon_{\rm Sel}$ is the 
product of trigger, reconstruction, filtering, likelihood 
and $M_{track}$ cut efficiency.
Apart from the last, which was evaluated only by Monte Carlo simulations, the other efficiencies were evaluated from unbiased 
samples of data with similar kinematics. The total selection efficiency is about $60\%$. }
\item{The counting rate measurement is norma\-lized to the
integrated luminosity $\int\mathcal{L}$ dt. At KLOE the luminosity
is measured using large angle Bhabha events 
($55^\circ<\theta_{+,-}<125^\circ$, $\sigma=425$nb). The precision of 
the measurement depends on evaluation of experimental efficiencies and acceptances, as well as on the theoretical knowledge of the process
\footnote{Berends Kleiss generator~\cite{Berends:fs} and 
BABAYAGA~\cite{CarloniCalame:2000pz}}. 
The very good agreement of the experimental distributions 
($\theta_{+,-}$, $E_{+,-}$) 
with the event generators  and a cross check with an independent 
counter based on $e^{+}e^{-} \rightarrow \gamma \gamma(\gamma)$ 
indicates that our precision is better than $1\%$.}
\end{itemize}

Fig.~\ref{fig:kloe} shows the number of $\pi\pi\gamma$ events 
selected inside the small angle acceptance cuts
at the end of the selection chain, for $100$ bins between $0.02$ GeV$^2$ and $1.02$ GeV$^2$.
This distribution comes from $\sim \,73 pb^{-1}$ of analyzed 
data, approximately $1/7$ of the total KLOE data sample of $\sim 500 \,pb^{-1}$. More than 1,0830,00 events have been selected, corresponding to  $~15,000$ events/$pb^{-1}$. The statistical error {\it per bin} is therefore on the level of few permil.
Even before unfolding the spectrum for the detector resolution effects, the $\rho-\omega$ interference can be clearly seen, this demonstrates the excellent momentum resolution of the KLOE drift chamber. 

\subsection{Extracting the pion form factor}
Neglecting FSR interference, the pion form factor can be extracted 
as a function of $M^2_{\pi\pi}$ from the observed $\ppg$ cross section $\sigma_{\pi\pi\gamma}$ as:
\begin{equation}
|F_\pi(M_{\pi\pi}^2)|^2 = \frac{\sigma_{\pi\pi\gamma}}
{\sigma_{\pi\pi\gamma}({F_\pi=1})}
\label{eq:pion}
\end{equation}
where $\sigma_{\pi\pi\gamma}({F_\pi=1})=H$
is the NLO cross section for $e^{+}e^{-} \rightarrow \ppg$ (only ISR) 
under the assumption of pointlike pions. We obtain the quantity $H$
bin-by-bin from the theoretical Monte Carlo generator PHOKHARA by setting $F_\pi=1$. 

The pion form factor extracted from eq.~\ref{eq:pion},
is shown in Fig.~\ref{fig:pion}. The spectrum 
can be divided into three regions with different errors: 
below $0.4\, GeV^2$ with an error of  5-10\%, between
$0.4\, GeV^2$ and $0.5\, GeV^2$, with an average error of 3\%
and above $0.5\, GeV^2$ with an 
average error of 2\%.  The errors are dominated
by the limited Monte Carlo statistics used in the
evaluation of the efficiencies and will be reduced with higher  MC statistics.

\subsection{Fit to the pion form factor}
A preliminary fit was applied to data using 
the K\"uhn-Santamaria parametrization for $F_\pi(Q^2)$~\cite{Kuhn:1990ad} (KS).
The mass and the width of the $\rho$ as well as $\alpha$ and $\beta$  
were free parameters of the fit while the other paramaters were kept fixed 
(to the  values of~\cite{Akhmetshin:2001ig}).

The results obtained from the fit are: $M_{\rho} = (772.6\pm0.5)\,MeV$, 
$\Gamma_{\rho} = (143.7\pm0.7)\,MeV$, 
$\alpha = (1.48\pm0.12)\cdot10^{-3}$, $\beta = -0.147\pm0.002$.
These values, even if preliminary, are in good agreement with 
the ones found by CMD-2
using a different parametrization (Gounaris-Sakurai). 
This agreement is confirmed  
in Fig.~\ref{fig:pion}, which shows 
the results of the two parametrizations (KLOE {\it blue line}, 
CMD-2 {\it red line}) superimposed to our experimental data. 
The discrepancy at low $M^2_{\pi\pi}$ is due to residual 
background events.

\section {Radiative return at BABAR}

Preliminary studies of the radiative return have been
performed at the PEP-II $b$-factory with BABAR data
~\cite{solodov01}. An event sample of 
$22$fb$^{-1}$ at $W=m(\Upsilon(4S))$ and continuum data have been 
analyzed so far, $\approx 1/4$ of the
total BABAR data. The event selection searches for exclusive hadronic 
final states accompanied by a hard photon ($1-9$ GeV).
It has been shown in ~\cite{eidelman99} that the acceptance of the BABAR 
detector for ISR-events is roughly $10-15\%$.  
Radiative dimuon events $e^+e^- \rightarrow \mu^+\mu^-\gamma$
have also been selected, which allows 
(for a specific final state $f$) a normalization of the hadronic 
cross section data:
\begin{equation}
\sigma_{f}({\rm M}_{\rm hadr}^2) =
\frac
{dN_{f\gamma}\cdot\epsilon_{\mu\mu}\cdot(1+\delta_{rad}^{\mu\mu})}
{dN_{\mu\mu\gamma}\cdot\epsilon_{f}\cdot(1+\delta_{rad}^{f})}
\cdot \sigma_{\mu+\mu-\gamma}
\end{equation}
where
$\epsilon_{\mu\mu}$ and $\epsilon_{f}$ are the detection
efficiencies and $\delta_{rad}^{\mu\mu}$, $\delta_{rad}^{f}$ are
final state radiative correction factors.
The radiative corrections to the initial state and the
acceptance for the ISR photon are
the same for $\mu\mu$ and $f$ and cancel in the ratio.

Preliminary results have been shown for the $\pi^+\pi^-$, $\pi^+\pi^-\pi^+\pi^-$ and $\pi^+\pi^-\pi^0\pi^0$. 
These distributions are very encouraging 
and show qualitatively a good agreement with the existing data.
The actual systematic errors are estimated to be at the few percent
level \footnote{Oliver Buchm\"uller, SLAC, private communication} and 
will be reduced with the ongoing analysis. 
Additional work is in progress for
the final states $K^+K^-$, $p \bar p$, $KK\pi^0$ and for higher 
multiplicities up to $7\pi$.

The BABAR results are of great importance for the energy 
region between $1.4$GeV and $2.0$ GeV, where the 
discrepancy between different experiments is in the order 
$15-20\%$ and where a measurement e.g. of the $4\pi$ channel 
of $\approx 1\%$ will have a big impact on $a_\mu^{\rm hadr}$
and $\alpha(M_{Z})$. The full energy range of interest can
be covered {\it in one experiment} with these data, avoiding
the usual normalization problems, when results from different experiments
are combined in the analysis of $a_\mu$ and $\alpha(M_{Z})$.

\section {Summary and outlook}

The status of the KLOE analysis on $e^{+}e^{-} \rightarrow \ppg$ and 
preliminary results obtained on the pion form factor were presented. 
A comparison of the pion form factor with the fitted parametrization 
obtained by the CMD-2 collaboration, shows a good preliminary agreement. 
Such a comparison, even at the given accuracy of $2-3\%$, has
become very important in light of the disagreement between
$\tau$ and $e^+e^-$ reported in the recent evaluation of $a_\mu$~\cite{Davier:2002dy}.

In order to improve the accuracy on  $a^{had}_{\mu}$ 
a final precision for this measurement below $1\%$ is needed. 
To reach this level of precision, a better understanding
of the systematic effects, both on the experimental and the 
theoretical side is mandatory. 
The intense theoretical work on radiative corrections from 
different groups and the preliminary results presented here by KLOE, are a promising indication for achieving such a challenging task.

The BABAR collaboration has presented preliminary 
distributions for the 
$2\pi$ and $4\pi$ final states using the 
radiative return from the $\Upsilon(4S)$ resonance. 
The results for the 4-pion state are 
competitive with the existing data. 
Improved results, both from KLOE and BABAR,
are expected in the following months. The preliminary results obtained 
so far by the two experiments, show that the radiative return is  
complementary to the standard energy scan and that it can produce
precise hadronic cross section data.


\begin{thebibliography}{9}

\bibitem{Davier:2002dy}
M.~Davier, S.~Eidelman, A.~H\"ocker and Z.~Zhang,
arXiv: hep-ph/0208177 and references there

\bibitem{Czarnecki:2001pv}
A.~Czarnecki and W.~J.~Marciano,
Phys.\ Rev.\ D {\bf 64} (2001) 013014

\bibitem{bnl}
G.~W.~Bennett {\it et al.}  [Muon g-2 Collaboration],
Phys. Rev. Lett.  {\bf 89} (2002) 101804

\bibitem{Davier:1998si}
M.~Davier and A.~H\"ocker,
Phys.\ Lett.\ B {\bf 435} (1998) 427

\bibitem{EidJeg95}
S.~Eidelman and F.~Jegerlehner,
Z.\ Phys.\ C {\bf 67} (1995) 585

\bibitem{Akhmetshin:2001ig}
R.~R.~Akhmetshin {\it et al.}  [CMD-2 Collaboration],
Phys.\ Lett.\ B {\bf 527} (2002) 161

\bibitem{bes}
J.Z.~Bai {\it et al.} [BES Collaboration],
Phys. \ Rev. \ Lett. \ B {513} (2001) 46

\bibitem{Spagnolo:1998mt}
S.~Spagnolo,
Eur.\ Phys.\ J.\ C {\bf 6} (1999) 637

\bibitem{Binner:1999bt}
S.~Binner, J.~H.~K\"uhn and K.~Melnikov,
Phys.\ Lett.\ B {\bf 459} (1999) 279

\bibitem{czyz02} 
H.~Czy{\.z}, these proceedings and references there

\bibitem{Hoefer:2001mx}
A.~H\"ofer, J.~Gluza and F.~Jegerlehner,
Eur.\ Phys.\ J.\ C {\bf 24} (2002) 51

\bibitem{Khoze:2002ix}
V.~A.~Khoze, M.~I.~Konchatnij, N.~P.~Merenkov, G.~Pancheri, L.~Trentadue and O.~N.~Shekhovtzova,
Eur.\ Phys.\ J.\ C {\bf 25} (2002) 199
and references there

\bibitem{Rodrigo:2001jr}
G.~Rodrigo, A.~Gehrmann-De Ridder, M.~Guilleaume and J.~H.~K\"uhn,
Eur.\ Phys.\ J.\ C {\bf 22} (2001) 81
[hep-ph/0106132].

\bibitem{Rodrigo:2001kf}
G.~Rodrigo, H.~Czy{\.z}, J.~H.~K{\"u}hn and M.~Szopa,
Eur.\ Phys.\ J.\ C {\bf 24} (2002) 71
[hep-ph/0112184].

\bibitem{Kuhn:2002xg}
J.~H.~K{\"u}hn and G.~Rodrigo,
Eur.\ Phys.\ J.\ C {\bf 25} (2002) 215
[hep-ph/0204283].

\bibitem{czyz01}
H.~Czy{\.z} and J.~H.~K\"uhn,
Eur.\ Phys.\ J.\ C {\bf 18} (2001) 497
%
\bibitem{Cataldi:1999dc}
G.~Cataldi, A.~Denig, W.~Kluge, S.~M\"uller and G.~Venanzoni,
Published in Frascati Physics Series (2000) 569-578

\bibitem{lp01}
A.~Aloisio {\it et al.}  [KLOE Collaboration],
Contributed paper to the 20th International Symposium on Lepton and Photon Interactions at High Energies, Rome, 23-28 July 2001

\bibitem{achim}
A.~Denig for the [KLOE Collaboration],
Proceedings of the 2001 $e^+ e^-$ Physics at intermediate energies 
workshop, SLAC/Stanford, April 30 - May 2 (2001), 
eConf: C010430 (2001) T07

\bibitem{dc}
M.~Adinolfi {\it et al.},
Nucl.\ Instrum.\ Meth.\ A {\bf 488} (2002) 51

\bibitem{emc}
M.~Adinolfi {\it et al.},
Nucl.\ Instrum.\ Meth.\ A {\bf 482} (2002) 364

\bibitem{Berends:fs}
F.~A.~Berends and R.~Kleiss,
Nucl.\ Phys.\ B {\bf 228} (1983) 537

\bibitem{CarloniCalame:2000pz}
C.~M.~Carloni Calame, C.~Lunardini, G.~Montagna, O.~Nicrosini and F.~Piccinini,
Nucl.\ Phys.\ B {\bf 584} (2000) 459

%
\bibitem{Kuhn:1990ad}
J.~H.~K\"uhn and A.~Santamaria,
Z.\ Phys.\ C {\bf 48} (1990) 445

\bibitem{solodov01}
E.~P.~Solodov
Proceedings of the 2001 $e^+ e^-$ Physics at intermediate energies 
workshop, SLAC/Stanford, April 30 - May 2 (2001), 
eConf: C010430 (2001) T03

\bibitem{eidelman99}
M.~Benayoun, S.~I.~Eidelman, V.~N.~Ivanchenko, Z.~K.~Silagadze, 
Modern Phys. Lett. A {14} 37 (1999) 2605

\end{thebibliography}
\end{document}